\documentclass[fleqn,usenatbib]{mnras}

\usepackage{newtxtext,newtxmath}
\usepackage[T1]{fontenc}
\usepackage{orcidlink}

\DeclareRobustCommand{\VAN}[3]{#2}
\let\VANthebibliography\thebibliography
\def\thebibliography{\DeclareRobustCommand{\VAN}[3]{##3}\VANthebibliography}

\usepackage{graphicx}	
\usepackage{amsmath}	
\usepackage{comment}
\usepackage{hyperref}

\title[Structure formation in hybrid MOND]{The formation of the first cosmic structures in hybrid Milgromian cosmology}

\author[T. Ondro \& N. Samaras]{
Tom\'{a}\v{s} Ondro$^{1,2}$$^{\orcidlink{0000-0001-5215-3066}}$\thanks{E-mail: tomas.ondro@mendelu.cz}
and Nikolaos Samaras$^{3,4}$$^{\orcidlink{0000-0001-8375-6652}}$\thanks{E-mail: nsamaras@uottawa.ca}
\\
$^{1}$Department of Technology and Automobile Transport, Faculty of AgriSciences, Mendel University in Brno, Zem\v{e}d\v{e}lsk\'{a} 1, 613 00 Brno, Czech Republic\\
$^{2}$Department of Applied Mathematics, VSB - Technical University of Ostrava, 17. listopadu 15/2172, 70800 Ostrava, Czech Republic\\
$^{3}$Astronomical Institute, Faculty of Mathematics and Physics, Charles University, V Hole\v{s}ovi\v{c}k\'ach2, CZ-180 00 Praha 8, Czech Republic\\
$^4$Department of Physics, University of Ottawa, Ottawa K1N 6N5, Canada
}

\date{Accepted XXX. Received YYY; in original form ZZZ}

\pubyear{\the\year{}}

\begin{document}
\label{firstpage}
\pagerange{\pageref{firstpage}--\pageref{lastpage}}
\maketitle

\begin{abstract}
Tackling cosmology with Milgromian Dynamics (MOND) has been a challenge ever since its theoretical inception. In a hybrid MOND scheme, a hypothesized light sterile neutrino of $\approx$ 11 eV/c\textsuperscript{2} rest-mass is able to provide realistic initial conditions, fitting the third peak of the \textit{Planck} CMB power spectrum as a Hot Dark Matter (HDM) component. At the same time, MOND governs the gravitational dynamics of density perturbations, while the background expansion follows the standard FLRW metric. We conduct and analyze hydrodynamical simulations of the fiducial $\Lambda$CDM model ($H_0 = 67.8$ km/s/Mpc, $\Omega_m \approx 0.31$, $\Omega_{\text{CDM}} \approx 0.26$), the progenitor $\nu$HDM model ($H_0 = 67.64$ km/s/Mpc, $\Omega_m = 0.3, \Omega_{\nu}\approx 0.26$), and its optimized version (opt-$\nu$HDM model, $H_0 \approx 55.6$ km/s/Mpc, $\Omega_m \approx 0.49$, $\Omega_{\nu} \approx 0.42$), utilizing the publicly available codes RAMSES and Phantom of RAMSES. We compare for the first time the total matter power spectrum $P(k)$ of the hydrodynamical simulations at various redshifts, examining the systematic differences of the three investigated cosmologies. The redshift evolution of the power spectrum of the $\Lambda$CDM model compared to the $\nu$HDM variants' counterpart offers important insight about the onset of structure formation. The combination of a light sterile neutrino and MOND gravity exceeds the $\Lambda$CDM model's matter $P(k)$ over all collapsing scales at late times ($z\lesssim1$). The density and velocity fields reveal a distinct difference between the underlying laws of gravity. Regardless of the modeled dependence in the jointly estimated data from \textit{Planck, DES Y1, SDSS, and eBOSS}, the $\nu$HDM hybrid models seem to produce an excess of power at associated resolution scales due to their cosmological design.
\end{abstract}


\section{Introduction}\label{sec:intro}

Over the past three decades, thanks to large galaxy redshift surveys, cosmology has been transformed into a quantitative branch of astrophysics. The extensive and high-precision observational legacy has enhanced our understanding about the distribution of matter on extragalactic scales. Extracting the encoded cosmological information enables us to constraint fundamental parameters and to examine the viability of various theoretical models. Projects such as the Baryon Oscillation Spectroscopic Survey \citep[BOSS]{2013AJ....145...10D}, being part of the Sloan Digital Sky Survey \citep[SDSS]{2013ApJ...768..117G}, probe the clustering of matter using 1.5 million luminous galaxies by measuring the Baryon Acoustic Oscillations (BAO). Built upon it, the extended Baryon Oscillation Spectroscopic Survey \citep[eBOSS]{2016AJ....151...44D} program, using the optical 2.5m Sloan Foundation Telescope at the Apache Point Observatory in New Mexico \citep{2006AJ....131.2332G}, expanded the selection of luminous galaxies in the redshift range $0.7<z<0.9$, investigating the history of dark energy.

The two-point correlation function $\xi(r)$ quantifies the spatial correlation of matter-density fluctuations at a separation $r$, relative to an unclustered distribution, while its Fourier transform, the matter power spectrum $P(k)$, describes the amplitude of these fluctuations as a function of wavenumber $k$. According to the conventional cosmological model, primordial density fluctuations began as (almost) scale-invariant and adiabatic in the early Universe \citep{1970PhRvD...1.2726H, 1972MNRAS.160P...1Z, 1970ApJ...162..815P} and resulted from an exponentially rapid expansion phase during cosmic inflation. The initial tiny perturbations are characterized by a power law, $P\left( k \right) \propto k^{n_s}$, $n_s \approx 1$ and $k$ being the comoving wavenumber. The amplitude grows as a function of time through gravitational effects, leading to non-linear evolution, giving birth to structures like galaxies or groups we observe today.

By measuring temperature anisotropies in the  $10^{-5}$ level of the non-uniform blackbody spectrum, all-sky surveys such as  \textit{WMAP} \citep{2013ApJS..208...20B} and \textit{Planck} \citep{2024A&A...682A..37T}  have captured the density inhomogeneities imprinted in the surface of the last scattering on behalf of the cosmic microwave background (CMB) radiation. The widely accepted $\Lambda$CDM cosmological model  \citep{1990Natur.348..705E, 1995Natur.377..600O} now relies heavily on the CMB power spectrum to explain the clustering of the large-scale structure (but see \citet{2025NuPhB101716931G, 2026eppg.confE.224K}).

Despite the successes, significant tensions against the concordance model persist. Matter bulk flows \citep{10.1093/mnras/stad1984} are found to be highly improbable in the $\Lambda$CDM model, as well as an immense under-dense region of $\approx 600$ Mpc \citep{2020MNRAS.499.2845H, 10.1093/mnras/stad3357, 2025MNRAS.536.3232M, 2026arXiv260203928B}. Claims against the statistically isotropic CMB, which is used as initial conditions for cosmological simulations, are raised as well \citep{2023arXiv231012859J}. 

The $\Lambda$CDM architecture relies primarily on the existence of Cold Dark Matter (CDM) particles and their early decoupling from the primordial plasma, so that vast gravitational potential wells (CDM halos) constitute the sites where the baryonic gas cools down and turns into stars, forming a galaxy. Nevertheless, its particle detection remains elusive \citep{9mff-p5k6} and astrophysical experiments dispute its presence \citep{2023eppg.confE.231K}.

On the other hand, Milgromian Dynamics, or Modified Newtonian Dynamics \citep[MOND]{1983ApJ...270..365M, 1983ApJ...270..371M, 1983ApJ...270..384M} was proposed as alternative to explain the flat galactic rotation curves without invoking the existence of unseen dark matter. The MOND phenomenology has undergone several tests across various scales. These tests range from wide binary stars  \citep{2024MNRAS.533..729H, 2025MNRAS.537.2925H, 2026arXiv260714450C} to star clusters \citep{2025A&A...693A.127P}, and they cover a broad mass spectrum, from ultra-faint dwarfs to groups and clusters \citep{2026ApJ..1001...65M}. In numerous instances, such as in galaxy clusters \citep{2026PhRvD.113d3027Z} and potentially the Bullet cluster \citep{6zrp-q7c4, 2026arXiv260410811H}, the assertions regarding the virial mass discrepancy \citep{Sanders_1999} have been mitigated, and a covariant formulation is already available \citep{PhysRevD.106.104041}; however, MOND cosmological frameworks often strive to replicate the concordance $\Lambda$CDM model, thereby ultimately encountering identical challenges on cosmological scales (missing satellites \citep{1999ApJ...522...82K}, the too-big-to-fail conundrum \citep{2011MNRAS.415L..40B}, and early galaxy formation as observed by JWST \citep{Naidu2026}). The MOND community continues to expand (the White Paper is forthcoming), however the foundational cosmology is still in a rudimentary phase \citep{1999PhLA..253..273M}. Since the devoted simulations of relativistic MOND theories like AeST \citep[Aether scalar-tensor,][]{2024PhRvD.110d4015B} and Khronon \citep{2024JCAP...11..040B} have not yet been achieved, Milgormian cosmology settles down in a less coherent but functional route.

The hybrid cosmologies exist between the two seemingly incompatible frameworks of the $\Lambda$CDM model and MOND \citep{2015CaJPh..93..250M}. The combination of a light sterile neutrino, which acts as a HDM particle, and Milgromian gravitation, provides a valuable approach for exploring cosmology, utilizing the strengths of both phenomenological models. The $\nu$HDM cosmological model originates from Sanders' study \citep{Sanders_1999}, who first identified the discrepancy between the estimated dynamical MOND mass and the observed mass of gas-rich galaxy clusters. While dark baryons \citep{2008NewAR..51..906M, 2024A&A...688A..78K} or stellar remnants \citep{2026PhRvD.113d3027Z} can help reconcile this discrepancy, sterile neutrinos also play a role in fitting the third peak of the CMB power spectrum \citep{Samaras2025} without disrupting the dominance of MOND at the scale of galaxies. $N$-body simulations \citep{2026arXiv260221975R} and hydrodynamical simulations \citep{Wittenburg2023, 2025arXiv251115325S} have addressed the subsequent structure formation; however, several questions remain unresolved. Specifically, MOND cosmology has been further supported by \citet{10.1046/j.1365-8711.1998.01459.x, 2001ApJ...560....1S}, who argued that the enhanced gravity compared to Newtonian gravity would facilitate the formation of bound objects especially in the early Universe. Nevertheless, direct comparisons on the matter power spectrum are currently absent from the literature. \citet{2025arXiv250700912B} have successfully reproduced the $\Lambda$CDM power spectrum, $P(k)$, using purely baryonic MOND approaches.

In this study, we perform hydrodynamical simulations and develop our own numerical techniques described in Section \ref{sec:simulations}. We evaluate the resulting power spectrum $P\left( k\right)$ from the numerical runs and compare it to existing observations (Section \ref{sec:analysis}). We present the corresponding findings in Section \ref{sec:results}, and in Section \ref{sec:conclusions}, we conclude our study.

\section{Simulations} \label{sec:simulations}

\subsection{\texorpdfstring{The $\nu$HDM cosmological model and its variant opt-$\nu$HDM}{The nuHDM cosmological model and its variant opt-nuHDM}}  \label{cosmo_model}

The $\nu$HDM model is based on a conservative ansatz. The Big Bang singularity, the cosmic inflation, and the expansion history, dictated by the Friedmann-Lema\^{\i}tre-Robertson-Walker (FLRW) metric, are kept in place as in the $\Lambda$CDM picture. The crucial distinction between the $\nu$HDM models and the fiducial $\Lambda$CDM one is the substitution of the CDM particle with a light, neutral, right-handed sterile neutrino ($m^{{\nu}\text{HDM}}_{\nu_{\text{s}}}=$ 11 eV/c\textsuperscript{2} or $m^{\text{opt-}{\nu}\text{HDM}}_{\nu_{\text{s}}} \approx$ 13 eV/c\textsuperscript{2}) as well as the MONDian gravity instead of the Newtonian one. Our hybrid scheme assumes that the cosmological background follows the General Relativity (GR) Friedmann equations \citep{2009MNRAS.394..527A, 2020MNRAS.499.2845H, Wittenburg2023}, while the gravitational dynamics recover the GR Poisson equation in the high-acceleration (Newtonian) branch of the quasi-static regime \citep{2007PhRvD..75d4017Z,Thomas2023}. At low accelerations, the quasi-static gravitational dynamics are described by a MOND-type modified Poisson equation sourced by density perturbations $\delta\rho=\rho-\bar\rho$ \citep{2001ApJ...560....1S,10.1046/j.1365-8711.2002.05235.x,10.1111/j.1365-2966.2008.13961.x,10.1111/j.1365-2966.2011.19321.x,Katz_2013,10.1093/mnras/stw1130,Wittenburg2023}. The generalized Einstein-Aether (GEA) theory of \citet{2007PhRvD..75d4017Z} provides a possible realization of these assumptions. For coefficients $c_1=-c_3$ and $c_2=0$, so that $c_1+3c_2+c_3=0$, and a free function satisfying $\mathcal{F}(0)=0$, the vector field leaves the homogeneous expansion unchanged (their sec.~V). In the static weak-field limit, they obtained a modified Poisson equation for a single gravitational potential, of the form proposed by \citet{1984ApJ...286....7B}, assuming that the spatial components of the vector field are of higher than first order in the perturbations (their sec.~III). For the particular free function chosen in their sec.~VI, this equation reproduces the `simple' MOND interpolating function. On an expanding background, \citet[][secs.~4.1--4.3]{Thomas2023} showed that, within the GEA models and approximations they considered, cosmological MOND behavior can arise when $c_1+3c_2+c_3=0$. For an appropriate choice of the free function, the quasi-static gravitational equations recover the GR Poisson equation in the Newtonian branch, including in the corresponding linear perturbative regime, and yield a MOND-type Poisson equation sourced by density perturbations in the deep-MOND regime. In contrast, for $c_1+3c_2+c_3\neq0$, their analysis does not produce MONDian behavior on cosmological scales.

The MOND-type equations derived in these GEA limits have the nonlinear modified-Poisson structure associated with AQUAL \citep{1984ApJ...286....7B}, whereas our simulations employ the quasi-linear formulation \citep[QUMOND;][]{10.1111/j.1365-2966.2009.16184.x}. The two formulations produce identical gravitational fields in configurations with one-dimensional symmetry, such as spherical systems, when their interpolating functions are appropriately related, but they are not generally equivalent for arbitrary three-dimensional mass distributions. Nevertheless, cosmological simulations using both formulations have demonstrated qualitatively similar structure-formation behavior relative to $\Lambda$CDM for the models and initial conditions considered \citep{10.1093/mnras/stw1130}. \citet[][sec.~4.3]{Thomas2023} also noted that existing MOND $N$-body codes can be used to investigate structure formation in suitable GEA cosmologies, provided that the matter content and simulation parameters are chosen consistently, but cautioned that the initial conditions used in previous studies may not correspond to those required by a fully consistent GEA cosmology. In the present work, we adopt GR-based linear transfer functions and initial conditions as a phenomenological approximation, motivated by the expectation that MOND modifications are small in the early Universe \citep[][sec.~3.1.3]{10.1046/j.1365-8711.1998.01459.x,2020MNRAS.499.2845H}. However, establishing their consistency with a particular GEA theory would require a full treatment of the linear cosmological perturbations, including the additional vector field and the sterile-neutrino component, which is beyond the scope of this work. Furthermore, the QUMOND formulation adopted here has not been derived directly from the particular relativistic GEA model considered above. Our simulations should therefore be regarded as a phenomenological study, motivated by the theoretical framework discussed above, rather than an exact realization of a specific relativistic theory.

Our results may nevertheless be relevant to relativistic completions admitting a similar cosmological MOND limit. Alternative theories, such as AeST \citep{2021PhRvL.127p1302S} and the Khronon theory \citep{2024JCAP...11..040B,2025arXiv250700912B}, contain additional gravitational degrees of freedom that can provide effective dark-matter-like contributions to cosmological evolution, including approximately dust-like behavior for suitable choices of the model functions and parameters, without requiring the sterile-neutrino component adopted here. These theories have different underlying cosmological dynamics and are not directly investigated in the present work.

We work with a particular family of $\nu$HDM variant models, derived from the optimal fit of the \textit{Planck} CMB power spectrum. Firstly, the $\nu$HDM model leaves the rest of the parameters identical to the $\Lambda$CDM model ones, except the $\Omega_{\text{CDM}}$, which is forced to zero, and that part of the matter budget is attributed to the $\Omega_{\nu} \approx 0.26$. Secondly, in the optimized version ($H_0\approx55.6$ km/s/Mpc, $\Omega_m \approx0.49$, $\Omega_{\nu} \approx0.42$), there is an important lag on the onset (confirmed by using CAMB) of the reionization era ($z^{\Lambda\text{CDM}}_{\text{reion}} = 7.686 = z^{\nu\text{HDM}}_{\text{reion}} \neq z^{\text{opt-}\nu\text{HDM}}_{\text{reion}} = 5.941$) and an enhancement of 0.8 Gyr in the age prediction ($t^{\Lambda \text{CDM}}_0 = 13.7 \text{ Gyr }= t^{\nu \text{HDM}}_0 \neq t^{\text{opt-}\nu \text{HDM}}_0 = 14.9$ Gyr\footnote{There is a difference of $\approx$ 0.3 Gyr in the age calculation between CAMB and the \citet{Samaras2025} value.}, due to inflation of the total matter budget.  For more detailed descriptions, we refer the reader to \citet{Samaras2025, 2025arXiv251115325S}, and Samaras \& Kroupa (2026) (under revision) concerning the functionalities of this particular hybrid framework. 

From the particle physics point of view, developments of a Lagrangian accommodating a sterile neutrino hypothetical particle \citep{2007JHEP...01..091A, SALVIO2015428} as a Neutrino Minimal Standard Model ($\nu$MSM) extension have been achieved too. Evidence for neutrino oscillations emanates from solar neutrinos \citep{1998ApJ...496..505C} or reactor experiments \citep{PhysRevLett.90.021802}, but a detection of a sterile neutrino via experiments like KATRIN \citep{Mohanty:2023wl} and NOvA \citep{PhysRevLett.134.081804} incorporates many experimental degeneracies (mass range, mixing angles, etc.) and model dependence (3 active massless neutrinos and a sterile one). We adopt the standard production mechanism by \citet{1994PhRvL..72...17D} via oscillations in the early Universe.

The main goal of this study is to explore the first emergent structures in the $\nu$HDM alternative models. MOND has long been proposed \citep{2008MNRAS.386.1588S} as a potential resolution for the required enhancement of gravity on large scales in order to explain observations that are presumably incompatible with a scenario based on the Newtonian gravitational collapse inside the dark matter halos (standard $\Lambda$CDM model). The sterile neutrino compromise was a necessary adjustment to fit the CMB power spectrum and a way to silence the early claims for a mismatch in the dynamical mass of galaxy clusters \citep{Sanders_1999}. The combination of a HDM particle and the baryonic gas collapsing under MONDian gravity added a complication that has been sparsely explored. The distinctive interplay between the MONDian baryons and the light sterile neutrino allows the exploration of the "Dark Sector" and it appears as a tempting candidate to many cosmological puzzles.

\begin{figure*}
    \centering
    \includegraphics[width=\linewidth]{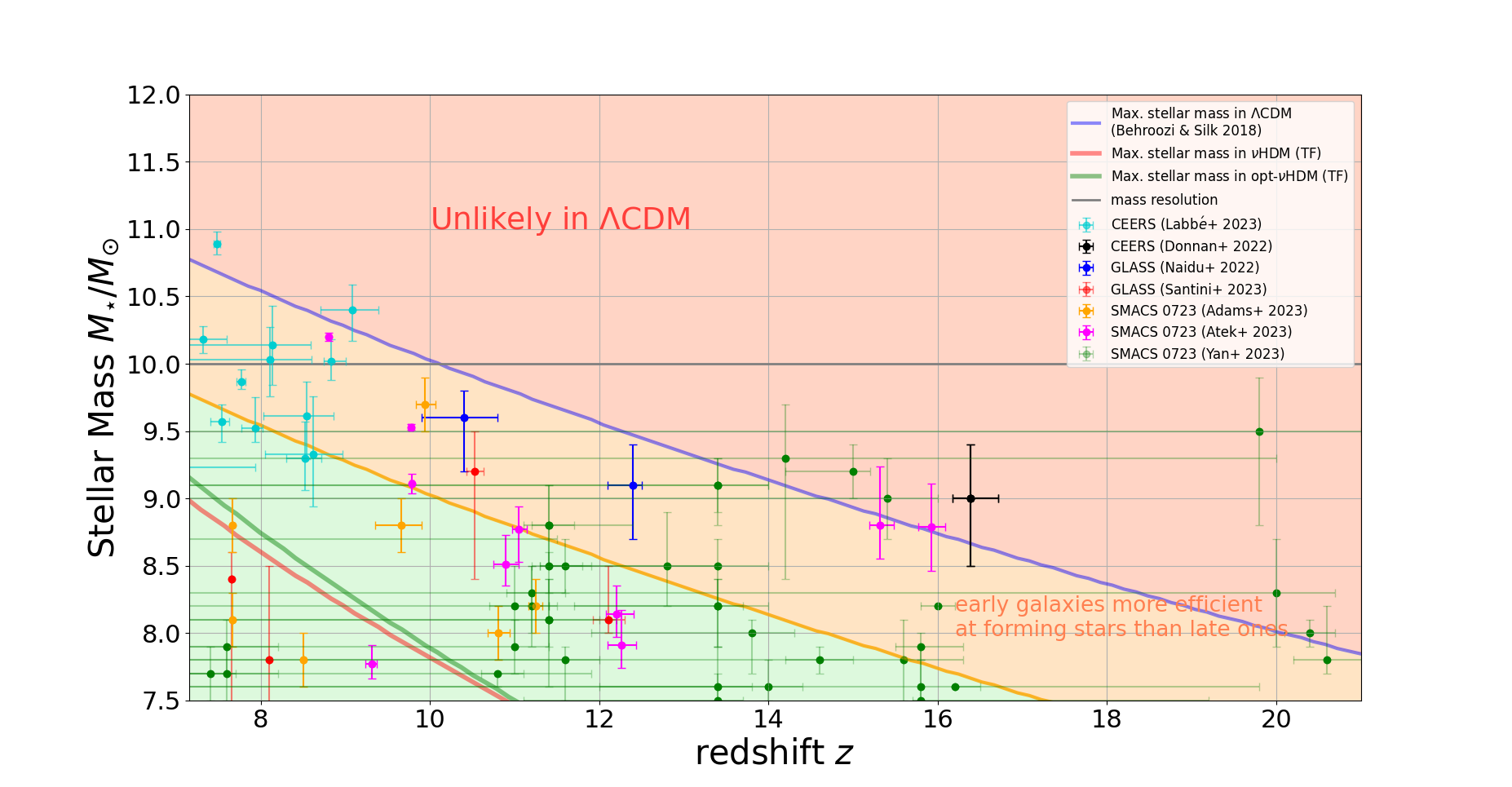}
    \caption{Linear Press-Schechter \citep{1974ApJ...187..425P} estimates of the maximum stellar mass $M_{\star}$ predicted in three different cosmological models, adapted from \citet[fig. 6]{2023mtsc.book...71L}. A rather non-plausible 100\% $\Lambda$CDM star-formation efficiency (SFE) (blue solid line) in the early universe ($7.5<z<22$) could reconcile the recent high-z observations (filled circles with asymmetric errorbars). Instead, a more reasonable 10\% SFE (orange line) would be in strong tension with the observational data. The 100\% SFE $\nu$HDM counterparts are shown in red ($\nu$HDM model) and green (opt-$\nu$HDM model) lines, in even stronger tension with the early galaxy formation. Nevertheless, the particle mass resolution in our PoR simulations is denoted by the gray horizontal line. Thus, even if the $\nu$HDM variant model does not predict, according to PS theory, such massive objects ($M_{\star}<10^{10} M_{\odot}$) in the early universe, they cannot be resolved anyway in a 200 $h^{-1}$Mpc simulation box with $256^3$ number of particles.
    The cosmic baryon fraction is assumed to be $f^{\Lambda\text{CDM}}_b = 0.157 \equiv f^{\nu\text{HDM}}_b \neq f^{\text{opt-}\nu\text{HDM}}_b = 0.15$ \citep{Planck2020} and the unitless Hubble constant today value is taken into account as $h^{\Lambda\text{CDM}} = 0.678\neq h^{\nu\text{HDM}} = 0.6764 \neq h^{\text{opt-}\nu\text{HDM}} = 0.556$. The cumulative number density is chosen to be $\Phi = 10^{-5}$Mpc\textsuperscript{-3}. For the standard $\Lambda$CDM model, an Eisenstein-Hu \citep{1998ApJ...496..605E} transfer function is used. The $\nu$HDM variant models are based on the transfer function derived by \citet{Samaras2025} at $z=199.0$.}
    \label{fig:mstar_z}
\end{figure*}

The James Webb Space Telescope (JWST) has uncovered unusually bright and rapidly growing galaxies in the early Universe \citep{Naidu2026, 2026arXiv260630787H}, contradicting most existing galaxy formation models. Using the {\sc hmf} package \citep[Halo Mass Function calculator]{2013A&C.....3...23M}, we first estimate the halo mass function with respect to redshift. Assuming a baryon fraction of $f^{\Lambda\text{CDM}}_b=0.157$ \citep{2020A&A...641A...6P}, $f^{\nu{\text{HDM}}}_b \approx 0.157$ and $f^{\text{opt-}\nu{\text{HDM}}}_b \approx 0.15$, we use the transfer functions from \citet{Samaras2025} for the hybrid models and convert the halo mass to the stellar mass one, comparing it to the $\Lambda$CDM model (Fig. \ref{fig:mstar_z}). The collection of high-z galaxy observational data is taken from \citet[fig. 6]{2023mtsc.book...71L}. Assuming a star-formation efficiency (SFE) of 100\% (blue line), 10\% (orange line), and any value below which is likely in the $\Lambda$CDM model (green shaded region) 
for the $\Lambda$CDM model, the shaded regions in Fig. \ref{fig:mstar_z} show various thresholds of estimated maximum stellar masses for a cumulative number density of 1 object in a volume of $10^{5}$ Mpc\textsuperscript{3}. Compared to the local Universe, the 100\% SFE is ten times more efficient. The ability of galaxies in the early $\Lambda$CDM universe to convert all of their gas into stars appears to have no clear explanation. If the survey volume was larger, the number density of galaxies would be smaller and in accord with the $\Lambda$CDM scenario based on abundance matching \citep{2018MNRAS.477.5382B}. Conversely, larger number densities are problematic for the $\Lambda$CDM model.

The halo mass function is approximated by a Press-Schechter \citep{1974ApJ...187..425P} function, which dictates a hierarchical structure formation scenario \citep{1978MNRAS.183..341W}. However, both hybrid MOND models are forming gravitationally bound objects in a top-down scheme because of the sterile neutrino free-streaming property, where the Zeldovich pancakes are fragmented to smaller bodies. For the $\nu$HDM variant models, the {\sc hmf } software allows the use of the transfer functions at $z=199.0$ from \citet{Samaras2025} as inputs. The late structure formation is evident for both hybrid models. The red line corresponds to the progenitor $\nu$HDM model and the green one to the opt-$\nu$HDM cosmology. The two lines are not parallel to each other (either to the $\Lambda$CDM line) because of the difference in the total matter budget ($\Omega^{\nu\text{HDM}}_m = 0.3 \neq \Omega^{\text{opt-}\nu\text{HDM}}_m \approx0.49$) and the Hubble constant ($H^{\nu\text{HDM}}_0=67.64 \text{ km/s/Mpc }\neq H^{\text{opt-}\nu\text{HDM}}_0 = 55.6$ km/s/Mpc). In early times ($z>10$), both hybrid models produce the same maximum stellar mass estimates, but in later times ($z\lesssim 8$), the opt-$\nu$HDM is expected to produce more massive stellar objects. Nevertheless, the resolution (horizontal gray line in Fig. \ref{fig:mstar_z}) of hydrodynamical simulations (Section \ref{sec:hydro_runs}) we have conducted is not sufficiently high to detect those stellar particles. None of the earlier studies by \citet{Wittenburg2023, 2026arXiv260221975R, 2025arXiv251115325S} focused on stars; therefore, this serves as further confirmation of the delayed formation of bound objects within this hybrid MOND framework.

\subsection{Initial conditions}\label{sec:ic}

The initial conditions (IC) were generated at $z = 199.0$ by jointly employing the {\sc camb} \citep{Lewis2000} and the {\sc music} \citep{Hahn2011} codes. The three cosmologies studied are flat ($\Omega_k = 0$), and their expansion histories conform to the FLRW metric. At high redshift ($z\approx200$)\footnote{According to \citet{Angulo2013} at higher redshifts ($z\approx200$), there remains a small contribution ($\approx 6 \%$) of the radiation energy density to the Friedmann equation, that we ignored in our simulations.}, the typical accelerations in the CMB frame exceed the MOND acceleration $a_0$. As a result, structure formation is expected to be minimally affected by MOND until $z\lesssim50$ \citep[sec 3.1.3]{2020MNRAS.499.2845H}. For the standard $\Lambda$CDM model, we utilize the cosmological parameters consistent with \citet{Planck2014}: $(\Omega_{\Lambda}, \Omega_{\rm m}, \Omega_{\rm b}, \sigma_{8}, n_{\rm s}, h, Y)= (0.692, 0.308, 0.0482, 0.829, 0.961,$ $ 0.678, 0.24)$. In the case of the $\nu$HDM and opt-$\nu$HDM frameworks, the cosmological parameters are sourced from \citet{Wittenburg2023} for $\nu$HDM and from \citet{Samaras2025} for opt-$\nu$HDM, respectively. \citet{2025arXiv251115325S} and Samaras \& Kroupa (2026, under revision) have outlined the analytical method for generating density and velocity fields in the early Universe regarding these hybrid MOND cosmologies. Therefore, we present a summary of the cosmological parameters in Table \ref{tab:cosmo_params}.

\begin{table}
    \centering
    \begin{tabular}{c c c c}
    \hline
    Model & $\nu$HDM & opt-$\nu$HDM\\
    \hline
        $\omega_{\nu} = \Omega_{\nu} h^{2}$ & 0.1205 & $0.1307 \pm 0.001$ \\
        $\omega_{\rm c} = \Omega_{\rm c} h^{2}$   & 0.000645 & 0.00001\\
        $\omega_{\rm b} = \Omega_{\rm b} h^{2}$ & 0.022383 & 0.022383\\
        $\Omega_{\rm m}$   & 0.3 & $0.495 \pm 0.009$  \\
        $H_{0} $ [km s$^{-1}$Mpc$^{-1}$]  & 67.64 & $55.64 \pm 0.32$ \\
        $A_{\rm PI}$  &  1.0 & $1.0\pm 0.0025$ \\
        $n_{s}$   & 0.966 & $0.872 \pm  0.0037$ \\
        $\tau$  & 0.0543 & $0.0375\pm 0.0069$ \\
        $\ln{(10^{10} A_{s})}$  & 3.0448 & $2.987\pm 0.015$ \\   
    \hline
    \end{tabular}
    \caption{Cosmological parameters of the hybrid cosmologies.}
    \label{tab:cosmo_params}
\end{table}



\subsection{Hydrodynamical simulation} \label{sec:hydro_runs}

We perform the $\Lambda$CDM simulations utilizing the Adaptive Mesh Refinement (AMR) code {\sc ramses} \citep{Teyssier2002} and the MONDian ones with the customized version of the {\sc ramses} code called {\sc phantom of ramses} \citep{Lughausen2015, Nagesh2021} (hereafter {\sc PoR}). We chose the minimal refinement parameter, which sets the size of the coarse grid to $l_{\rm min} = 8$, whereas the maximum level of refinement was set to $l_{\rm max} = 9$ \footnote{The size of grid is calculated as $n_{x} = 2^{l}$.}. One should note that we used relatively low resolution; however, simulations suites for cosmological studies like Quijote \citep{Villaescusa2020} or Aemulus \citep{DeRose2019} have smaller or comparable resolutions to our simulations. The runs commence at $z=199.0$ for a  200 $h^{-1}$Mpc box, and star-formation is not activated. Importantly, we do not follow the \citet{Wittenburg2023} formalism in identifying bound structures in order to understand the binding procedure (Section \ref{sec:find_bound_objects}).

\section{Analysis of simulations} \label{sec:analysis}

\subsection{Matter power spectra}
We calculate the total matter power spectrum by taking the fast Fourier transform of the dimensionless field $\delta_{\rm m}$, which is usually called the density contrast as
\begin{equation}
    \delta_{\rm m} = \frac{\rho_{\rm m} - \overline{\rho}_{\rm m}}{\overline{\rho}_{\rm m}}.
\end{equation}
Considering that the total matter is composed of collisionless matter (dark matter in case of $\Lambda$CDM and sterile neutrinos in $\nu$HDM variant models) and baryons, we firstly prepare a 3D density array by interpolating the particles onto the $N = 256^{3}$ uniform grid via cloud-in-cell (CIC) method. Then, we sum those components as 
\begin{equation}
    \rho_{\rm m} = \rho_{\rm DM} + \rho_{\rm gas}
\end{equation}
to have one total matter density cube $\rho_{\rm m}$.

\begin{figure*}
    \centering
    \includegraphics[width=0.85\linewidth]{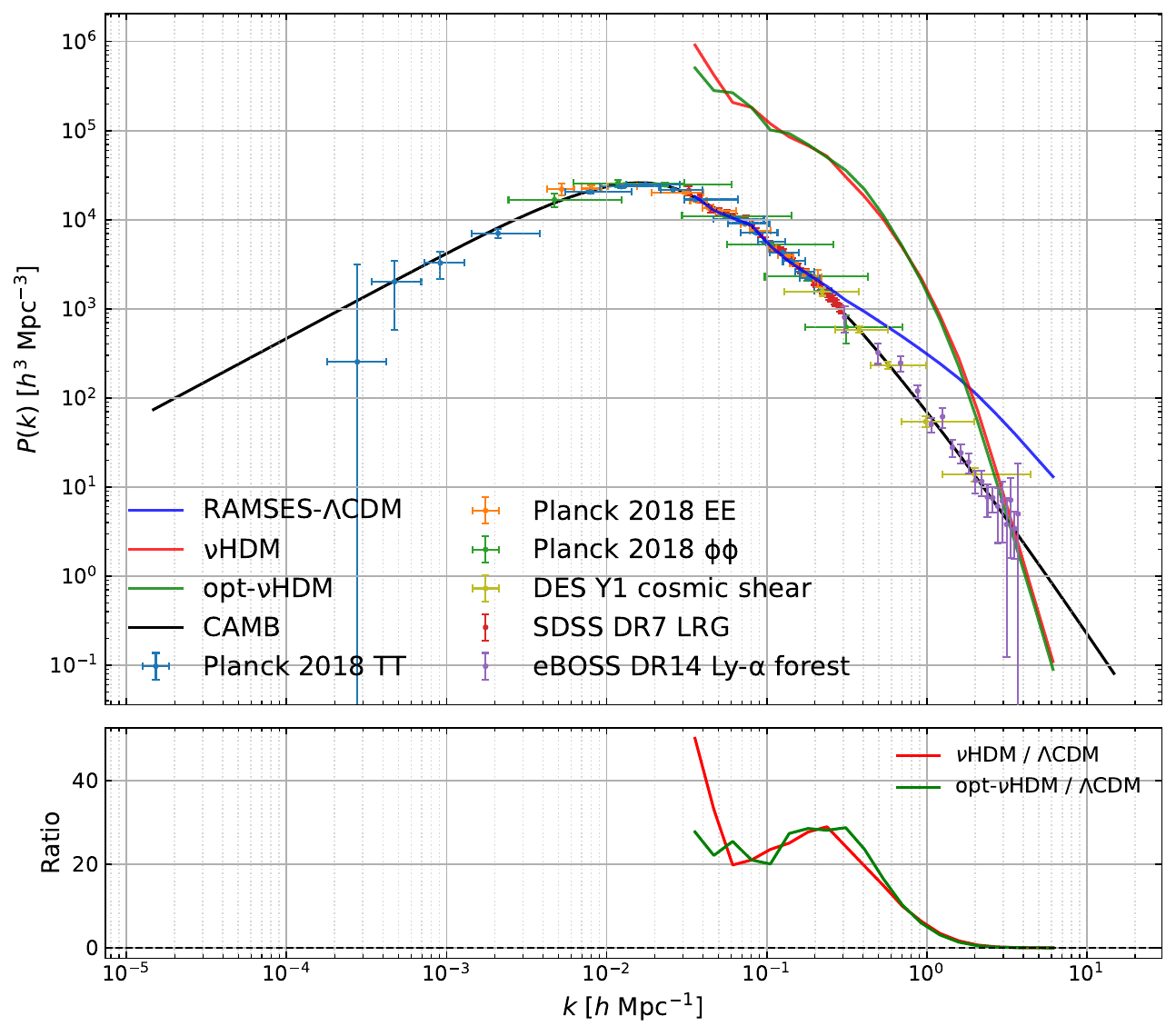}
    \caption{Inferences of the 3D linear total matter power spectrum at $z=0$ from 
    \textit{Planck} \citep{Planck2020} CMB data on the largest scales, \textit{SDSS} \citep{2010MNRAS.404...60R} galaxy clustering on intermediate scales, and \textit{eBOSS} \citep{2018ApJS..235...42A}, SDSS Ly-$\alpha$ clustering and \textit{DES Y1} \citep{2018PhRvD..98d3528T} cosmic shear data on the smallest scales. The solid blue, red and green lines correspond to the total matter power spectra from simulations evolved with {\sc ramses} and {\sc PoR} codes for the $\Lambda$CDM, the $\nu$HDM and the opt-$\nu$HDM models, respectively. The solid black line represents the total matter power spectrum as computed by the {\sc camb} Boltzmann solver. Both hybrid MOND models over-predict the amount of power at small scales $k>10^{-2}$ $h$ Mpc\textsuperscript{-1}, because of the combination of the MONDian gravitational effects and the sterile neutrino clustering. The overabundance of power is similar for both models.}
    \label{fig:pk_all}
\end{figure*}

\subsection{Estimation of the density field on a grid}
We construct 2D projections of matter over-density $\Delta_i$ at the grid point $i$ given by:
\begin{equation}
    \Delta_{i} = \frac{\rho_i}{\overline{\rho}} \text{, with }  \overline{\rho} = \frac{1}{N}\sum^{N}_{i = 1} \rho_{i}
\end{equation}
where $\rho_i$ is density at the grid point $i$ and $\overline{\rho}$ is the average density. For particle distributions (CDM for the $\Lambda$CDM model and sterile neutrinos for the hybrid MOND models), we prepare a density cube by interpolating the particles onto the $N = 256^{3}$ uniform grid via the CIC method \citep{1988csup.book.....H, 2022LRCA....8....1A}, which is a common numerical method for particle-grid interpolation. 
Note that the choice of interpolation scheme was a compromise between accuracy and computational expense. In case of the hydro variables (e.g. pressure and density), we calculated the Cartesian cube using the internal {\sc ramses} routine {\sc amr2cube}. The structure and evolution of cosmic web can be characterized by the probability distribution function (PDF) of the matter density field \citep{Einasto2021}.

\subsection{Estimation of the temperature field on a grid}
Due to our computational resources, we do not include radiative heating and cooling, and therefore we do not have information about the ionization fractions. However each {\sc ramses} and {\sc PoR} output snapshot contains the information about the thermodynamic pressure $P_{\rm cell}$. The gas internal energy can be derived from pressure as
\begin{equation}
u = \frac{P_{\rm cell}}{(\gamma - 1) \times \rho_{\rm gas}},
\end{equation}
where $\gamma = 5/3$ for monoatomic gas composition. Then the gas temperature can be derived from internal energy as
\begin{equation}
T = (\gamma - 1) \times \frac{\mu \times m_{\rm p}}{k_{\rm B}} \times u,
\end{equation}
where $k_{\rm B}$ is the Boltzmann constant, $\mu$ denotes the mean molecular weight and $m_p$ is the
mass of a proton. Note that this derivation requires the mean molecular mass of the gas to be known. This depends on the ionization state, which we do not store. However, we can calculate it using following very crude approximations
\begin{equation}
\mu = 4 / (4 - 3 \times Y), \qquad {\rm for}\,T < 10^{4}\,{\rm K}
\label{Eq:mol_weight1}
\end{equation}
and
\begin{equation}
\mu = 4 / (8 - 5 \times Y), \qquad {\rm for}\,T \geq 10^{4}\,{\rm K}
\label{Eq:mol_weight2}
\end{equation}
where $Y = 0.24$ is the helium mass fraction \citep{Gaikwad2017, 2021A&A...649A.117G}.

\subsection{Finding gravitationally bound structures} \label{sec:find_bound_objects}

The cosmogonic action of the $\nu$HDM variant models is intricate due to the cooperation of the sterile neutrinos and the MONDian gravity. The cosmic web displays a well-defined filamentary structure, with large galaxy clusters situated at the nodes. The overall abundance aligns with observations (Samaras \& Kroupa 2026, under revision); however, earlier studies \citep{10.1111/j.1365-2966.2011.19321.x, Katz_2013} indicated late structure formation and the emergence of excessively massive clusters, with total masses around $M_{\text{tot}}\approx 10^{17}M_{\odot}$.

Unlike earlier studies, we do not use Amiga Halo Finder \citep[AHF]{Knollmann2009} to detect gravitationally bound objects. Bias against low-mass structures is reported \citep[sec 2.5]{Wittenburg2023}. 
Furthermore, according to the Halo-Finder Comparison Project \citep{10.1093/mnras/sts173},  agreement was obtained between different codes regarding the total, dark and stellar components of the bound objects. However, the diffuse gas content of the haloes shows great disparity, especially for low-mass satellite galaxies. Hence, we decide to adopt two different halo finding techniques\footnote{Analysis was performed using the {\sc yt\_astro\_analysis} extension
\citep{yt.astro.analysis} of the {\sc yt} analysis toolkit \citep{Turk2011}.}:
\begin{itemize}
    \item The Friends-of-Friends (FoF) method \citep{Efstathiou1985,Turk2011}: the basic principle of this method is that we firstly calculate the average inter-particle spacing from the total number of particles and the volume of the region. Then, the pairs of particles closer than the inter-particle distance (compared to overall average) $l_{\rm FoF} = 0.2$ are linked together. Particles can be paired with more than one other particle. The final groups are formed the networks of particles linked together by friends.
    \item The HOP algorithm \citep{Eisenstein1998,Turk2011}: We begin by estimating the local density at each particle using a smoothing kernel. Then the algorithm builds chains of linked particles by hopping from one particle to its densest neighbor. A particle which is its own densest neighbor is the end of the chain. All chains that share the same densest particle are grouped together. Groups are included, linked together, or discarded depending on the over-density threshold parameter $\delta = 160$.
\end{itemize}
Note that the chosen values of $l_{\rm FoF}$ and $\delta$ in case of FoF and HOP method, respectively, are typical for such applications (see \cite{Turk2011, yt.astro.analysis}).

\section{Results and discussion}\label{sec:results}
\subsection{Matter power spectra}
We highlight the total matter power spectra of the three cosmologies at $z=0$, comparing them with a compilation of data\footnote{We make use of the open access Jupyter notebook \url{https://github.com/marius311/mpk\_compilation} from \citet{Chabanier2019}.} from \textit{Planck} \citep{Planck2020} - \textit{DES Y1} \citep{2018PhRvD..98d3528T} - \textit{SDSS} \citep{2010MNRAS.404...60R} - \textit{eBOSS} \citep{2018ApJS..235...42A} in Fig. \ref{fig:pk_all}. This analysis underscores the enhanced power of the hybrid MOND cosmogony. The simulated surplus of power arises from the interplay between sterile neutrinos and MOND gravity. The overall shape of the power spectrum \( P(k) \) deviates from that of the $\Lambda$CDM model across all corresponding scales. In both the \(\nu\)HDM and opt-\(\nu\)HDM models, the power spectrum exhibits a "knee" in the range of \( 10^{-1} \, h \, \text{Mpc}^{-1} < k < 10^0 \, h \, \text{Mpc}^{-1} \), followed by a steeper decline at smaller scales (\( k > 10^0 \, h \, \text{Mpc}^{-1} \)). At even smaller scales (\( k \gtrsim 10^{0.5} \, h \, \text{Mpc}^{-1} \)), the hybrid cosmologies show a significant suppression of power, attributed to the free-streaming of sterile neutrinos and the subsequent erasure of the evolving cosmic structure. The $\Lambda$CDM model, which is based on a collisionless cold dark matter (CDM) component, exhibits significantly more power at the scales of \(1 \, h \, \text{Mpc}^{-1} < k < 6 \, h \, \text{Mpc}^{-1}\). This overabundance of power— approximately a factor $\approx 20 - 30$ — occurs at the scales where structure formation takes place, specifically in the range \(10^{-2} \, h \, \text{Mpc}^{-1} < k < 10^0 \, h \, \text{Mpc}^{-1}\). This phenomenon helps to account for the presence of massive bound objects and the relative emptiness of void regions in hybrid scenarios. However, it is important to note that the observations are fitted to $P\left( k \right)$ templates of the $\Lambda$CDM model, making them not entirely independent \citep{Tegmark2002}.

\begin{figure*}
    \centering
    \includegraphics[width=\linewidth]{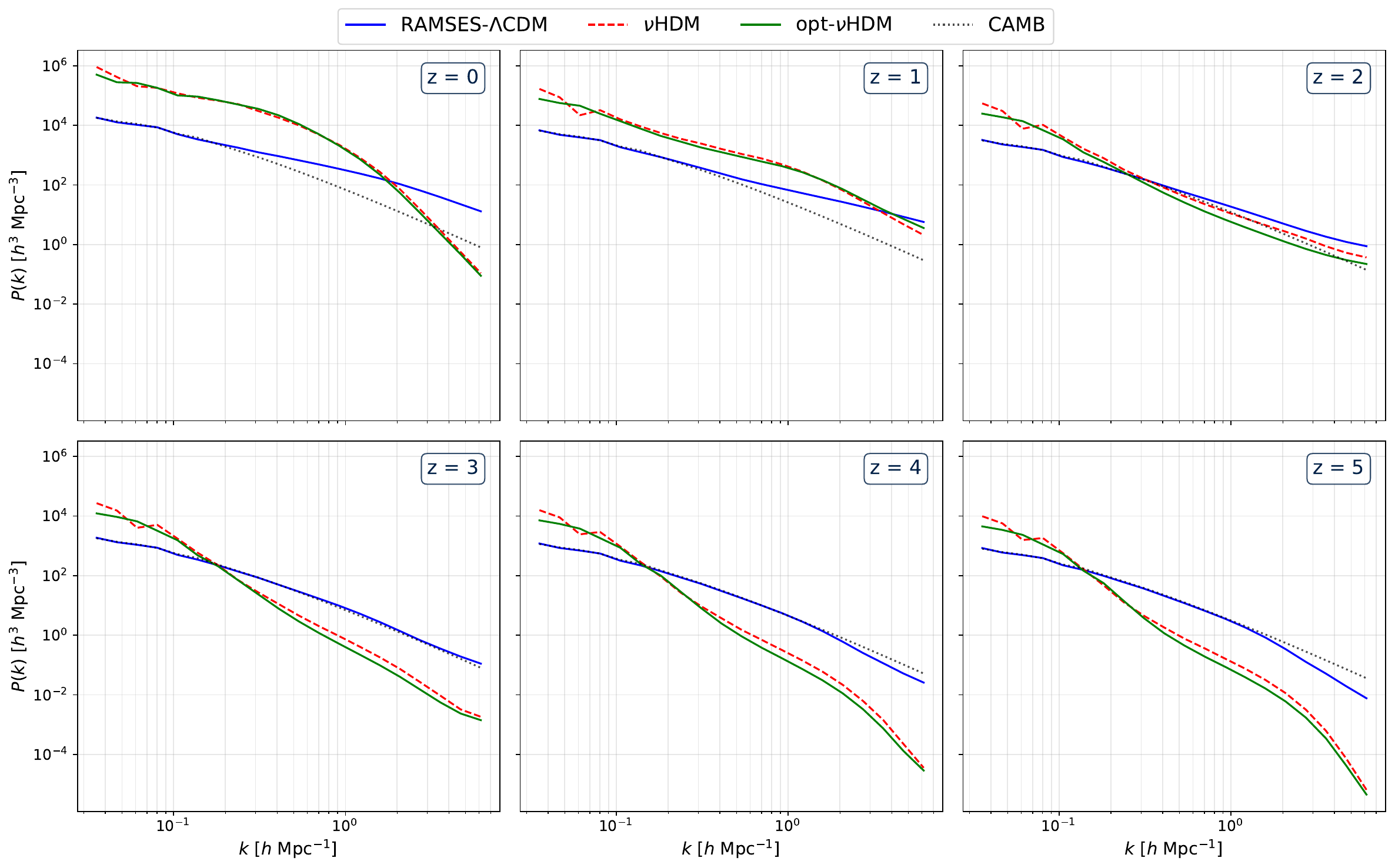}
    \caption{Matter power spectra from simulation evolved with {\sc ramses} and {\sc PoR} codes. We also plotted the matter power spectrum as computed by the {\sc camb} Boltzmann solver \citep{Lewis2000}.}
    \label{fig:PS}
\end{figure*}

The matter power spectra from hydrodynamical simulations evolved with {\sc ramses} and {\sc PoR} codes are illustrated in Fig. \ref{fig:PS}. In this plot, we use blue to represent the $\Lambda$CDM model, red for the $\nu$HDM model, and green for the opt-$\nu$HDM model, focusing on the redshift range of $0<z<5$. At early times ($2<z<5$), a distinct pattern emerges: at very large scales $k\lesssim10^{-1}h$ Mpc\textsuperscript{-1}, the hybrid MOND models exhibit more power than the $\Lambda$CDM model due to their top-down structure formation approach. The emerging objects form through monolithic collapse and subsequent fragmentation. At lower scales ($k \gtrsim 10^{0}h$ Mpc\textsuperscript{-1}), the effects of sterile neutrinos become apparent, as structure formation experiences significant delays compared to standard cosmology. The free-streaming nature of the right-handed HDM component suppresses the enhanced gravitational effects of MOND, resulting in the absence of bound structures at redshifts higher than 3. This suggests that the hybrid MOND cosmologies could prove to be unproductive of high-z galaxies.

On the other hand, for later times ($0 \leq z\leq2$), the pattern of $P \left( k \right)$ differs in the upper row of Fig. \ref{fig:PS}. The hybrid MOND cosmologies continue to exhibit more power on large scales $k\lesssim10^{-1}h$ Mpc\textsuperscript{-1}. However, on smaller scales, the evolution of cosmic structure becomes more complex. The three distinct cosmologies show roughly the same power at $z=2$. As time progresses, fragmentation gives way to gravitational collapse under MOND gravity. Sterile neutrinos, which are no longer relativistic and can now cluster, work together with baryonic gas to begin forming bound objects. This process creates over-densities that increase the power compared to the $\Lambda$CDM model. It is rather unclear what the power spectrum would be in the absence of sterile neutrinos, but their combination with MOND exceed the $\Lambda$CDM model estimates at low-$z$. At very small scales ($k \approx 10^{0.5}h$ Mpc\textsuperscript{-1}), the sterile neutrino effects become more apparent, suppressing the power to a lower than the $\Lambda$CDM level. This is a late-time effect of their free-streaming inside the simulation volume. Note that we also plot the redshift-dependent total matter power spectra in Appendix \ref{Sec:Appendix_A} for all the studied cosmologies.

\subsection{Density Projections}
\begin{figure*}
    \centering
    \includegraphics[width=\linewidth]{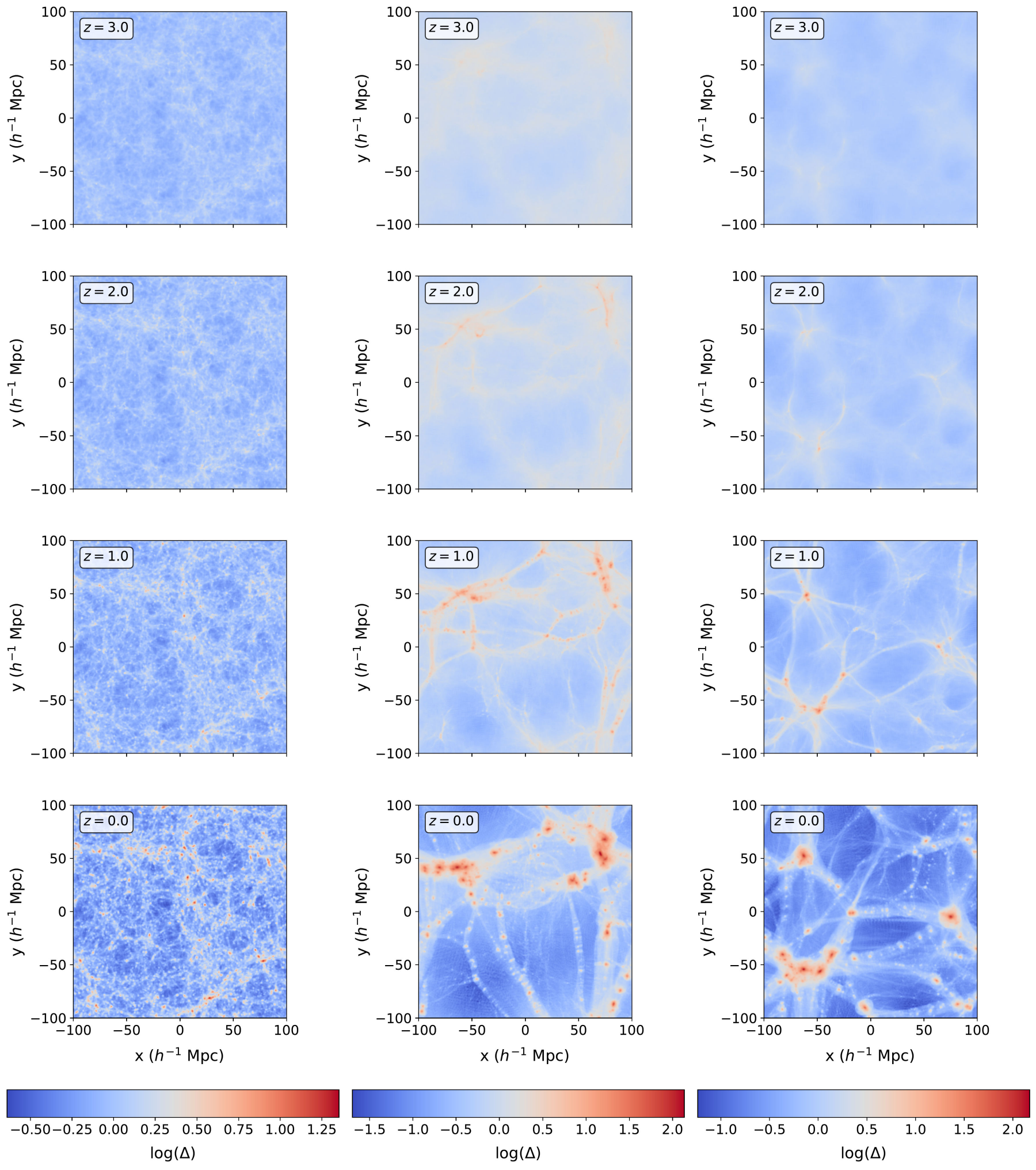}
    \caption{Comparison of simulations evolved with {\sc ramses} and {\sc PoR} codes. From top to bottom the rows correspond to projection of over-densities at different epochs of the cold dark matter (CDM) component (left column) for the standard $\Lambda$CDM model, as well as for sterile neutrinos in the hybrid models (the $\nu$HDM model in the middle and the opt-$\nu$HDM at the right).}
    \label{fig:part_projection}
\end{figure*}

Fig. \ref{fig:part_projection} presents projections of the cold dark matter (CDM) component for the standard $\Lambda$CDM model, projections of sterile neutrinos over-densities for the hybrid models. The left column represents the standard $\Lambda$CDM cosmology, the middle column depicts the $\nu$HDM model, and the right column illustrates the opt-$\nu$HDM cosmology. Cosmic evolution progresses downwards to lower redshifts. A visual examination of the snapshots at $z=3$ indicates that the fiducial model is more fine-grained compared to the coarser hybrid counterparts. At early times, the $\nu$HDM model demonstrates the highest level of homogeneity, displaying minimal over-density values. The opt-$\nu$HDM model appears to exhibit the largest regions of emptiness. As time advances, the $\Lambda$CDM cosmology becomes increasingly detailed, revealing numerous micro-structures. In contrast, the hybrid MOND models develop distinct structures relatively early. At $z=2$, the $\nu$HDM over-densities are widespread throughout the simulation volume, indicating a potential for significant cosmogonic activity. The opt-$\nu$HDM model at intermediate redshifts ($1\lesssim z \lesssim 2$) appears more homogeneous than its predecessor, despite having a higher matter budget, with $\Omega^{\nu \text{HDM}}_m = 0.3$ compared to $\Omega^{\text{opt-}\nu\text{HDM}}_m = 0.49$. At $z=0$, the $\Lambda$CDM model exhibits a neat and smooth development of the cosmic web, matching large-scale observations \citep{Jones2015}. The sharp hybrid cosmogonies lead to the formation of massive bound objects and a distinct cosmic web, characterized by long, notable filaments and smaller galactic systems arranged along the spines. Additionally, the extended under-dense regions observed in the opt-$\nu$HDM simulation at $z=0$ align with theoretical predictions of MOND \citep[sec. 9]{2012LRR....15...10F}. Future research will explore the alignment of galactic spins within the filamentary sheets of MOND simulations.

\begin{figure*}
    \centering
    \includegraphics[width=\linewidth]{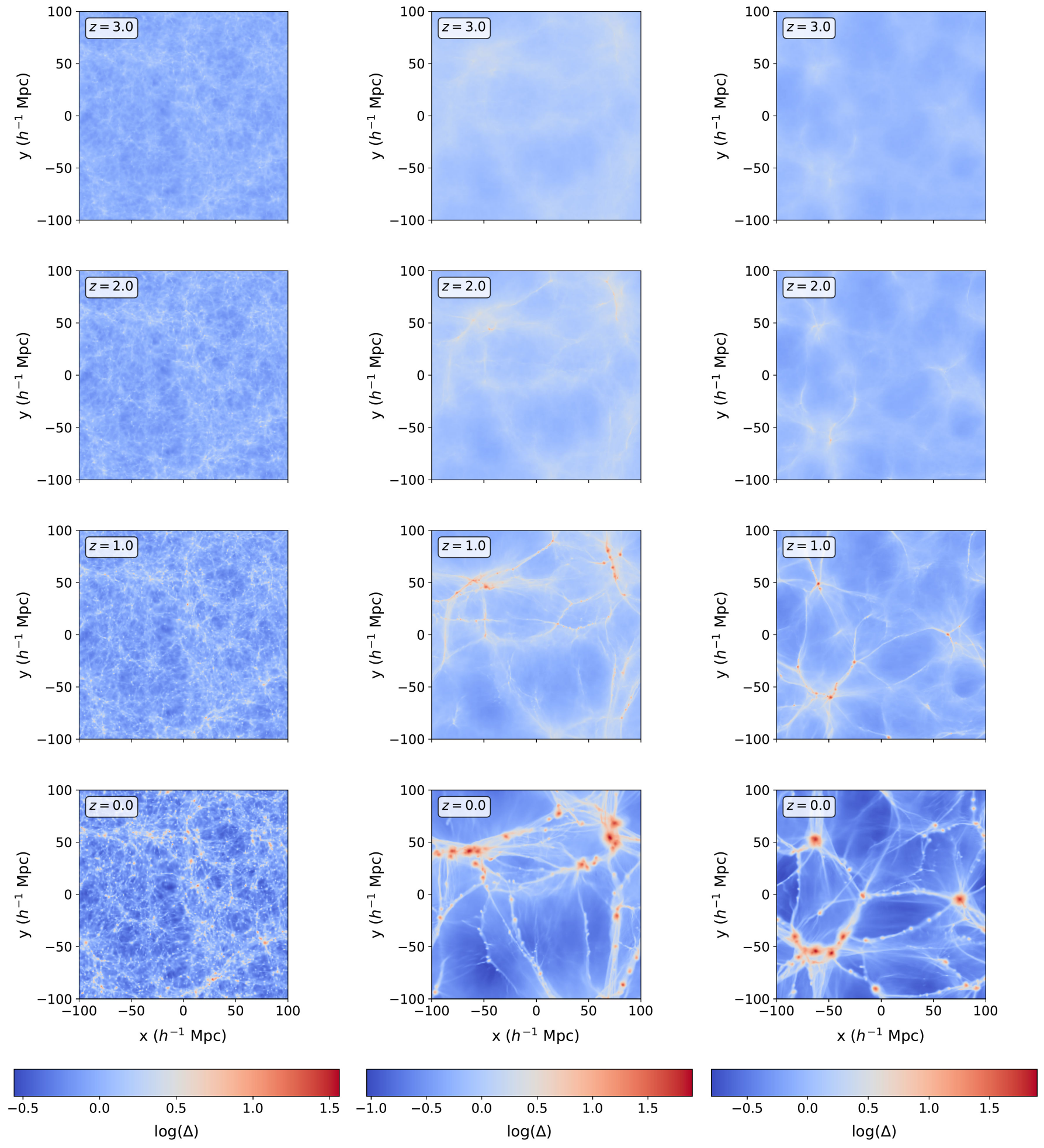}
    \caption{Same as Fig. \ref{fig:part_projection} but for projections of gas over-densities for the three studied cosmologies at different epochs.}
    \label{fig:gas_projection}
\end{figure*}

Furthermore, Fig. \ref{fig:gas_projection} illustrates the over-densities of baryonic matter as a function of redshift. In general, baryons tend to follow the distribution of dark matter, regardless of the type. The three cosmological models examined demonstrate significantly different patterns of gaseous evolution. The $\Lambda$CDM model generates structure on small scales, primarily through the merging process. On large scales, the structure in the $\Lambda$CDM model appears fairly uniform at redshifts $z\lesssim2$. The cosmic baryonic web develops gradually into larger structures, characterized by small voids and less prominent filaments. Similar to the distribution of sterile neutrinos, the emergence of cosmic bodies in the MONDian gaseous model is significantly delayed compared to the $\Lambda$CDM model. Fragmentation of large clouds takes place around $z\approx 2$ (see also Section \ref{time_evol}). Hybrid MOND models develop massive bound structures earlier than those in the fiducial cosmology, particularly evident in the panels at $z=1$. Additionally, the overall scale of homogeneity is achieved at higher scales in the MOND models, which lack the finer and more intricate structure characteristic of the $\Lambda$CDM cosmogony. The opt-$\nu$HDM model (shown in the right column) appears to be the most affected by sterile neutrino streaming, as indicated by its increased sterile neutrino budget, with $\Omega^{\text{opt-}\nu{\text{HDM}}}_{\nu} \approx 0.42$ compared to $\Omega^{\nu\text{HDM}}_{\nu} \approx 0.26$. The baryonic gas in the hybrid MONDian model appears to be similar, with only a few specific locations where matter accumulates, leaving the rest of the cosmological model quiet and empty compared to the $\Lambda$CDM cosmic web, with extended filaments.

%

\subsection{Temperature Projections}
\begin{figure*}
    \centering
    \includegraphics[width=\linewidth]{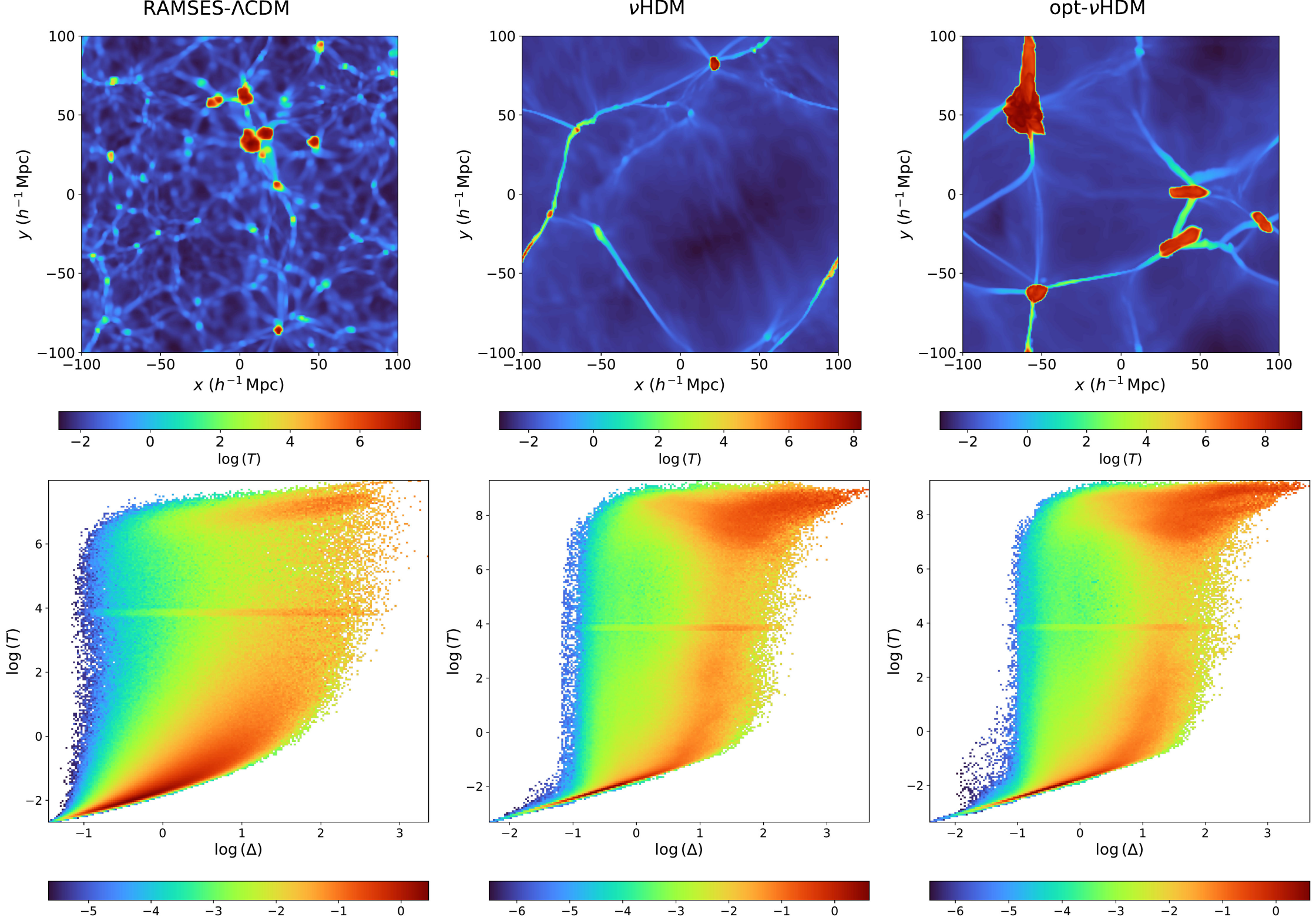}
    \caption{Upper panel: the calculated gas temperature $T$ at $z = 0$. The color scheme is such that the red and the blue color corresponds to highest temperature and lowest temperature regions, respectively. Slices from a simulation box in the first row have width $\approx 0.8\,h^{-1}$ Mpc. 
    Bottom panel: Distribution of gas cells in the $T - \Delta$ plane at $z = 0$. The color scheme represents density of points in logarithmic scale. Note that the artifact (horizontal line) around $T \approx 10^{4}$ K is caused by temperature calculation method, see Eqs. (\ref{Eq:mol_weight1}) and (\ref{Eq:mol_weight2}).}
    \label{fig:temp_dens}
\end{figure*}

We calculate the gas temperature and the gas over-densities in the $T-\Delta$ (log) plane (Fig. \ref{fig:temp_dens}). The three examined cosmologies display distinct temperature maps (upper panel of Fig. \ref{fig:temp_dens}). The $\Lambda$CDM model (left column) features numerous dispersed temperature peaks located at the nodes of cosmic filaments, indicating the presence of massive clusters of galactic systems. The temperature range is relatively narrow compared to that of the hybrid MOND models ($-2.5 \lesssim \log{(T^{\Lambda \text{CDM}} / \text{K})} \lesssim 7.5$). The $\nu$HDM model (middle panel) is the least dense of the three models, featuring a few isolated regions with higher temperatures. The filamentary structure appears less extensive compared to that of the $\Lambda$CDM model. In contrast, the opt-$\nu$HDM model (right column) exhibits long, thick regions of elevated temperature. Additionally, in several instances, the temperature maps are not oriented along the typical axis of the filaments.
The highly over-dense regions exhibit higher temperatures due to hydrodynamical processes; we will refer to this heating as shock heating resulting from structure formation. However, it is important to note that this numerical setup does not encompass all relevant physics, including the photo-heating of low-density gas by the ultraviolet background (UVB) to higher temperatures, either any cooling processes \citep{Ondro_2025}. Simulations that incorporate additional heating and cooling processes \citep{Hui1997} indicate that temperature and density maintain a reasonably tight correlation for mildly non-linear densities, specifically for $\Delta \leq 10$ \citep{Gaikwad2017,Gaikwad2018}. This tight correlation is also evident in the $\nu$HDM and opt-$\nu$HDM models. Therefore, we intend to investigate the properties of the gas which surrounds cosmic filaments in the context of the $\nu$HDM and opt-$\nu$HDM models in future research.

The lower panels of Fig. \ref{fig:temp_dens} show the dispersion of temperature values regarding the three cosmological models. The standard cosmology develops its structure in a hierarchical manner, resulting in a limited range of density contrast on the x-axis. In contrast, the hybrid MOND cosmologies, which combine sterile neutrinos with the MONDian enhancement of gravity, permit a broader spectrum of temperature and density inhomogeneities. 

\subsection{Evolution of bound objects} \label{time_evol}
\begin{figure*}
    \centering
    \includegraphics[width=\linewidth]{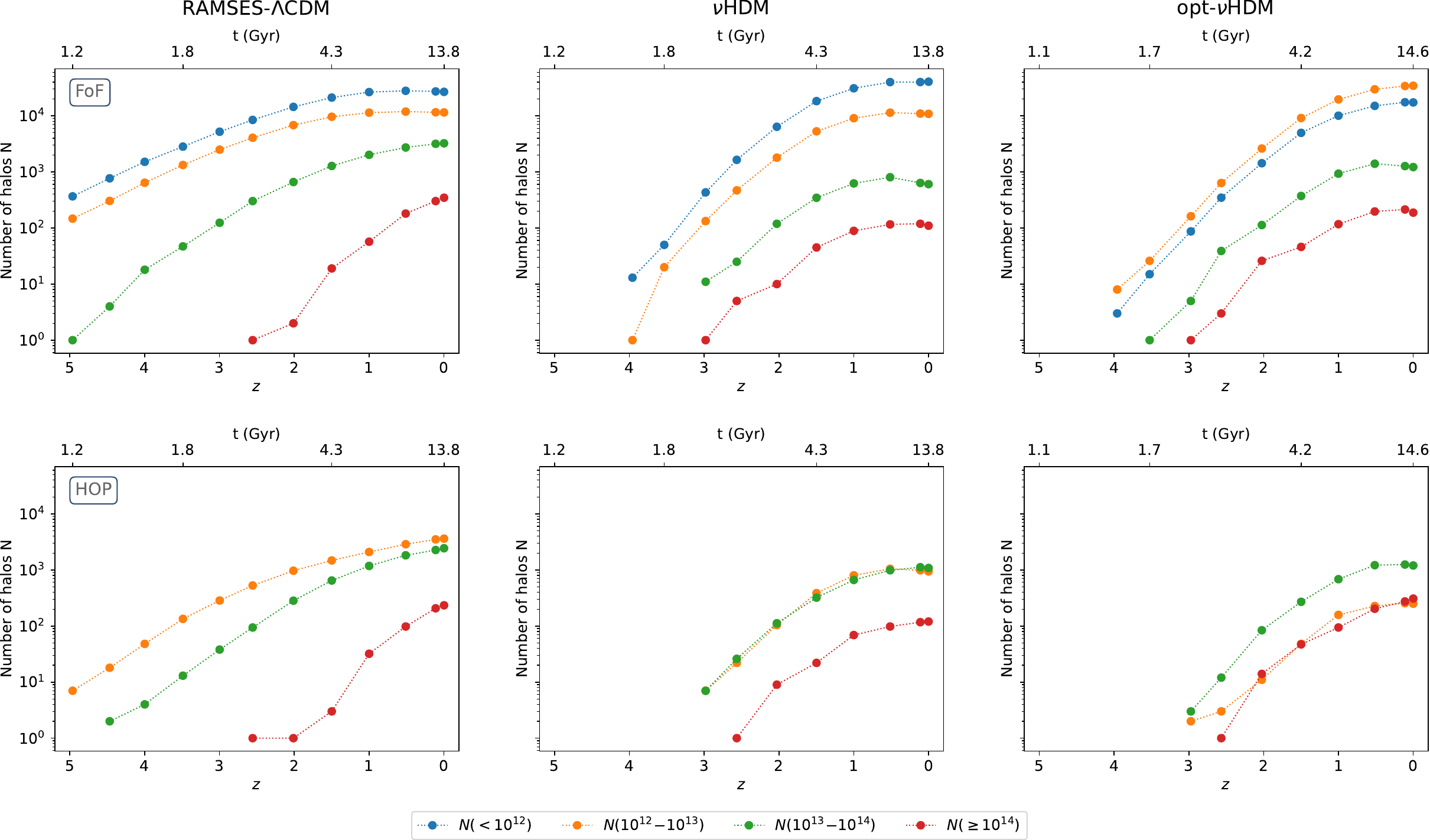}
    \caption{Evolution of number of bound structures from simulation evolved with {\sc ramses} (left column) and {\sc PoR} (2nd and 3rd column) codes. The results in first and second row correspond to the FoF and HOP halo finding method, respectively.}
    \label{fig:dm_halos}
\end{figure*}

\citet{Wittenburg2023, 2025arXiv251115325S} have detected objects at higher redshifts, $z \approx 4.5$ as spherical over-densities. Contrary to previous studies, we investigate the time evolution of the number of bound objects, as shown in Fig. \ref{fig:dm_halos} applying different numerical methods. It is important to note that our resolution was insufficient for identifying subhaloes; therefore, only individual haloes are analyzed. We utilize two distinct algorithms: the FoF in the upper panels and the HOP one in the lower panels. At the left of Fig. \ref{fig:dm_halos} is the $\Lambda$CDM model; the middle panel displays the $\nu$HDM model, while the right panel corresponds to the opt-$\nu$HDM model.  The $\Lambda$CDM model exhibits a greater number of low-mass objects compared to the hybrid models, which aligns with expectations based on the hierarchical nature of structure formation. The hybrid MOND models exhibit slow growth in a top-down scenario; however, after \( z \approx 2 \), galaxy formation becomes quite intense for the hybrid MOND scenarios, resulting in a shallower distribution of objects. When the sterile neutrinos cease to suppress power, the growth of structure becomes very energetic. Between the two hybrid models, the FoF algorithm indicates a lower number of low-mass bound objects for the opt-$\nu$HDM model, while the counts of massive clusters are similar.

The HOP technique indicates that low-mass objects ($<10^{12}M_{\odot}$) are absent in all three cosmogonic scenarios. The $\Lambda$CDM model reveals an earlier emergence of massive objects at around $z\approx5$, accompanied by a sharp increase in their quantity at lower redshifts. In contrast, the hybrid MOND models generate massive objects rapidly, and their quantities remain stable throughout cosmic evolution, resulting in a shallower distribution. In the low-resolution setup, none of the three cosmological models examined succeed in forming objects early enough to correspond with JWST observations \citep{Yan_2023, Harikane_2025, Naidu2026}. Future research will focus on increasing the particle count and enhancing resolution to further explore high-redshift objects.

Last, we report the most massive halos in the simulations. Based on the FoF algorithm:
\begin{itemize}
    \item \textit{max} $\text{M}_{\text{Tot}}$\textsuperscript{$\Lambda$CDM}$ = 2.25 \times 10^{15}M_{\odot}$
    \item \textit{max} $\text{M}_{\text{Tot}}$\textsuperscript{$\nu$HDM}$ = 7.6 \times 10^{16} M_{\odot}$
    \item \textit{max} $\text{M}_{\text{Tot}}$\textsuperscript{opt-$\nu$HDM}$ = 1.07 \times 10^{17}M_{\odot}$
\end{itemize} 

Based on the HOP technique:
\begin{itemize}
    \item \textit{max} $\text{M}_{\text{Tot}}$\textsuperscript{$\Lambda$CDM}$ = 1.75 \times 10^{15}M_{\odot}$
    \item \textit{max} $\text{M}_{\text{Tot}}$\textsuperscript{$\nu$HDM}$ = 5.6 \times 10^{16} M_{\odot}$
    \item \textit{max} $\text{M}_{\text{Tot}}$\textsuperscript{opt-$\nu$HDM}$ = 7.0 \times 10^{16}M_{\odot}$
\end{itemize} 

The mass estimates of the most massive bound structure are enhanced when using the FoF technique. We have revisited the well-known issue of the $\nu$HDM model, which produces unrealistically massive clusters \citep{Katz_2013, Wittenburg2023}. However, \citet[chap. 3]{2025arXiv251115325S} noted that by employing more complicated star-formation recipes and incorporating supernova feedback, the total mass of the most massive halos decreases by $\approx 30\%$.
%

\subsection{\texorpdfstring{Over-density \& Velocity distributions at $z=2$}{Over-density \& Velocity distributions at z=2}} \label{vel_distrib}
This section connects the previous one to the velocity distribution of three cosmological models by examining the probability density function (PDF) of the gas over-densities at $z=2$ for the three studied cosmologies collected from smooth coherent filaments. 

\begin{figure*}
    \centering
    \includegraphics[width=\linewidth]{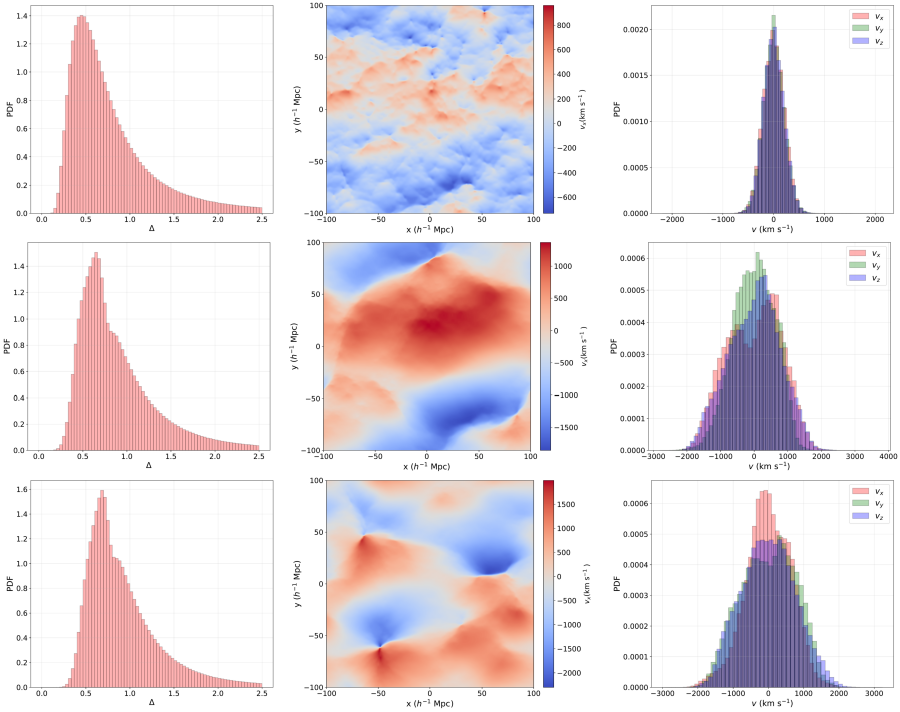}
    \caption{Results from simulations at $z = 2$ for $\Lambda$CDM (top), $\nu$HDM (middle) and opt-$\nu$HDM (bottom) models. Left-hand, middle and right-hand panels in each row show PDF of over-densities, (snapshot slice) velocity component ($v_{x}$) along x-axis and PDF of (cube) velocities, respectively. Slices from a simulation box in the middle columns have width $\approx \,0.8\,h^{-1}$ Mpc.}
    \label{fig:dens_vel}
\end{figure*}

The $\Lambda$CDM model exhibits an approximately log-normal distribution \citep{Kofman1994}, with its maximum at $\Delta_{\rm max}^{\Lambda{\rm CDM}} \approx 0.45$. In contrast, we identify a bimodal distribution for the hybrid MOND models: in both models, the first peak is at $\Delta \approx 0.65$--0.7 and the second at $\Delta \approx 0.85$, where it appears as a shoulder in the histograms of Fig.~\ref{fig:dens_vel}. All three PDFs are shown for $0 < \Delta < 2.5$ with the same binning. This may indicate the co-existence of sterile neutrinos and MONDian gravity.

\citet[fig. 4]{Katz_2013} study the galaxy cluster bulk flows resulting from $N$-body MOND simulations. They discovered significant MOND velocities. In our analysis, we confirm their findings by comparing them to the bulk velocities of the $\Lambda$CDM model, as shown in the middle and right panels of Fig. \ref{fig:dens_vel}. The first row represents the $\Lambda$CDM model, the middle row corresponds to the $\nu$HDM model, and the third row illustrates the opt-$\nu$HDM model. The velocity contrasts for the hybrid MOND models display distinct characteristics. The middle panel of Fig. \ref{fig:dens_vel} indicates that the velocity behavior (taken from a $\approx \,0.8\,h^{-1}$ Mpc slice from a 200 $h^{-1}$Mpc simulation box along the x-axis ($v_{x}$)) of the three cosmologies mirrors that of the density distributions shown in Fig. \ref{fig:gas_projection}. Similarly to the density projection, the velocity projection for the hybrid MOND models reveals extended high-velocity fields adjacent to low-velocity regions. The velocity field of the $\Lambda$CDM model appears more homogeneous, with less pronounced peaks in velocity and a broader range of velocity contrasts. In contrast, MOND is predicted to generate higher velocities than the $\Lambda$CDM model, which raises concerns regarding consistency with standard cosmology \citep{Tully_2023, 10.1093/mnras/stad1984}. The outflows can be associated with a large local void \citep{2020MNRAS.499.2845H, 10.1093/mnras/stad3357}. This void creates a potential hill, leading to galaxies flowing away from it and potentially addressing the Hubble tension \citep{2024arXiv241000804B, 2025PDU....4901965D}.

We examine each velocity component and present $v_x, v_y, v_z$ in the right panel of Fig. \ref{fig:dens_vel}. The Gaussian $\Lambda$CDM probability density function (PDF) distribution is consistent across all $v_i$ components. In contrast, the PDF for the hybrid models is significantly smaller and demonstrates anisotropic behavior. It remains to be determined whether this discrepancy is an apparent phenomenon or a prediction of Modified Newtonian Dynamics (MOND). The tails of the hybrid model's distribution exhibit a significantly wider range of values ($-2500\lesssim v^{\nu{\text{HDM}}}\text{[km/s]}\lesssim 2500$), which aligns with the expectations from MOND boost gravity. This discrepancy, in physical terms, results in the gas being drawn into the developing large-scale structure, thereby aiding in the clearing of voids.


\section{Conclusions} \label{sec:conclusions}
We have conducted hydrodynamical simulations based on the standard model of cosmology, $\Lambda$CDM, with parameters $H_0 = 67.8$ km/s/Mpc, $\Omega_m \approx 0.31$, and $\Omega_{\text{CDM}} \approx 0.26$. We compared these simulations with MONDian hybrid models, specifically the $\nu$HDM, which has parameters $H_0 = 67.64$ km/s/Mpc, $\Omega_m = 0.3$, and $\Omega_{\nu} \approx 0.26$, as well as the opt-$\nu$HDM model, characterized by $H_0 \approx 55.6$ km/s/Mpc, $\Omega_m \approx 0.49$, and $\Omega_{\nu} \approx 0.42$. The overall conclusion of this work is that sterile neutrinos present significant challenges for the hybrid MOND models. Although they provide a means to fit realistic initial conditions, such as the \textit{Planck} CMB power spectrum, they also impose limitations on the formation of the first bound structures (Fig. \ref{fig:mstar_z}) in cosmic evolution. The free-streaming phenomenon suppresses ongoing galaxy formation, acting as a hindrance to the enhanced gravitational effects predicted by MOND. The combined impact of these factors results in an excess of overall power across collapse scales, leading to an overestimation compared to the $\Lambda$CDM power spectrum, $P(k)$ (Fig. \ref{fig:pk_all}).

While this issue alone is not problematic, the observational data from the total matter power spectrum (Fig. \ref{fig:pk_all}) derived from \textit{Planck}, \textit{SDSS}, \textit{DES Y1} and \textit{eBOSS} are all model-dependent, similar to the \textit{Planck} CMB temperature fluctuation power spectrum \citep[sec. 5.5]{Samaras2025}. Nonetheless, a significant concern regarding the hybrid models under investigation is the delayed emergence of bound objects (Fig. \ref{fig:dm_halos}). The recent discoveries of high-redshift galaxies by JWST \citep{Harikane_2025, Naidu2026} raise important questions for any coherent cosmological framework, although the resolution of our simulations is not sufficient to search for low-mass early Universe structures. Additionally, the sterile neutrino serves to counterbalance the gravitational effects enhanced by MOND. More broadly, the various $\nu$HDM models remain viable contenders due to the limitations in resolution present in this analysis. The density and velocity fields in the hybrid MOND models are significantly more coarse-grained compared to the finer-grained fields in the $\Lambda$CDM models. Additionally, the temperature and over-density ranges in the $\nu$HDM variant models are broader, indicating an enhancement of the gravitational fields due to MOND. The hybrid models' total matter power spectrum at $z\approx2$ surpasses the predicted power of the $\Lambda$CDM model at large scales. In light of this, we analyze each of the velocity components, revealing an anisotropy. We remind the reader that more advanced hybrid simulations are available and appear to refine the estimates of the most massive hybrid objects at $z=0$. Future research will focus on star-formation techniques, proper photo-heating (by UV background) and resolution effects.

\section*{Acknowledgments}
The authors would like to express their gratitude to the Czechoslovakian student conference series (BEZOVEC), where they met and initiated their collaboration. The authors would also like to thank Tom Z{\l}o{\'s}nik, Bharat Ratra, Chan-Gyung Park and Peter Laursen for the fruitful discussions. NS received support from the Charles University Grant Agency (GAUK) - 94224, as well as from the Ostrava Supercomputing Center for the project OPEN-32-11 \#2885, where the $\nu$HDM simulations were conducted. NS received support from the Dynamix computer system at the Argelander Institute for Astronomy, Bonn University, as well as from the SPODYR group. Additionally, NS benefited from assistance through the Deutscher Akademischer Austauschdienst (DAAD) Bonn-Prague exchange program. Currently, NS is employed at the University of Ottawa in Ottawa, Ontario, Canada.

\section*{Data Availability}
We acknowledge the use of the Planck Legacy Archive. Planck (\url{http://www.esa.int/Planck}) is a European Space Agency (ESA) science mission with instruments and contributions directly funded by ESA Member States, NASA, and Canada. The \href{https://classic.sdss.org/home.php}{Sloan Digital Sky Survey (SDSS)} is a redshift survey using a dedicated 2.5-m wide-angle optical telescope at Apache Point Observatory in New Mexico, United States. We make use of  the Year 1 data release of the \href{https://www.darkenergysurvey.org/des-year-1-cosmology-results-papers/}{Dark Energy Survey (DES)}. \href{https://www.sdss4.org/surveys/eboss/}{eBOSS} is Spectroscopic Survey studying the Baryon Acoustic Oscillations. For reproducing this work, one will have to visit \href{https://github.com/NickSam121/nuHDM-cosmo-sims}{NS's GitHub page} and \href{https://github.com/TOndro90/Cosmo-analysis}{TO's page}.

\bibliographystyle{mnras}
\bibliography{BIBLIO} 

\appendix
\section{\texorpdfstring{Redshift evolution of $P\left( \lowercase{k} \right)$}{Redshift evolution of P(k)}} \label{Sec:Appendix_A}
\begin{figure*}
    \centering
    \includegraphics[width=\linewidth]{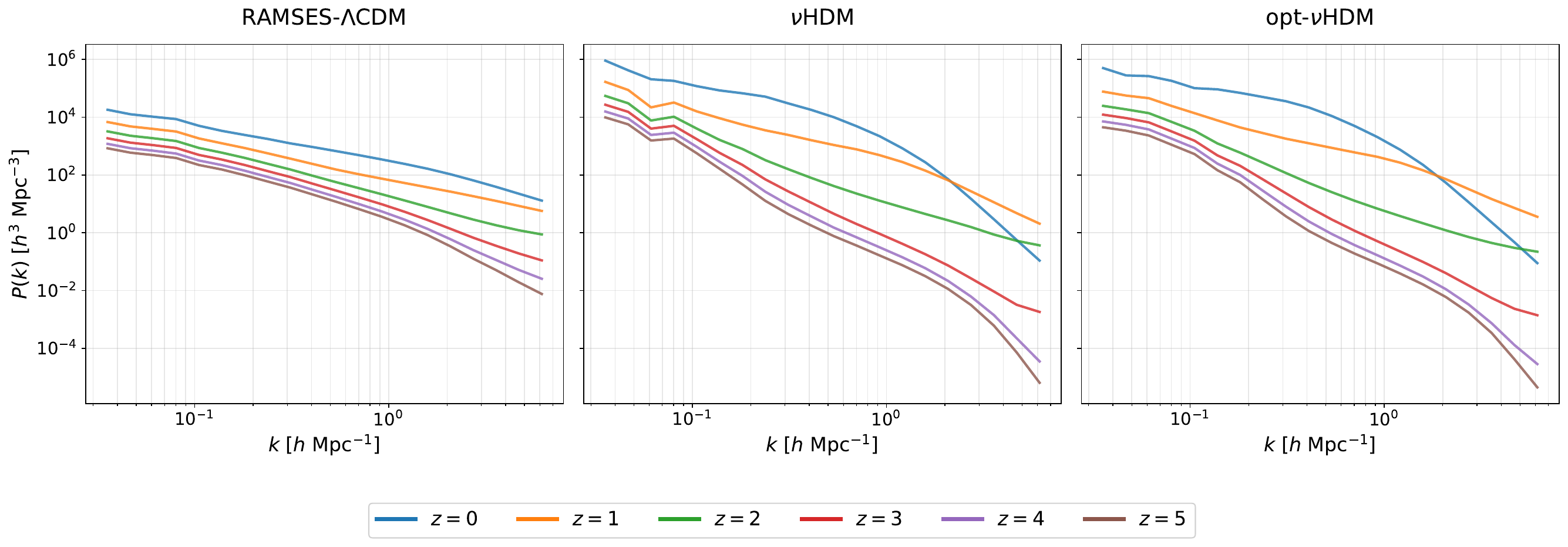}
    \caption{The redshift evolution of the total matter power spectra for the three studied cosmologies.}
    \label{fig:PS_3model}
\end{figure*}

The evolution of the total matter power spectrum is shown in Fig. \ref{fig:PS_3model}. The hybrid scenarios surpass the $\Lambda$CDM model in power at very large scales ($k < 10^{-1}h \text{ Mpc\textsuperscript{-1}}$) throughout all phases of structure formation. In late times ($z<2$), the total matter power spectra of the hybrid cosmologies depend on the interplay between MOND gravity and sterile neutrinos. Except for the very late times ($z=0$), the power systematically increases for both the $\nu$HDM model and the opt-$\nu$HDM model. It is only at the final snapshot ($z=0$) that the sterile neutrinos significantly suppress the power at scales $k\gtrapprox10^{0.5}h$ Mpc\textsuperscript{-1}. The cause of this effect—whether it stems from numerical resolution, free-streaming activities of sterile neutrinos, or a combination of both—will be investigated through simulations with varying box sizes. Nonetheless, this effect is particularly pronounced in the hybrid cosmologies, as the $\Lambda$CDM model consistently enhances the matter power spectrum $P\left( k \right)$ as redshift \( z \) decreases.


\bsp	
\label{lastpage}
\end{document}